\documentclass[review]{elsarticle}

\usepackage{xcolor}

\journal{Arxiv.org}

\usepackage{hyperref}

\begin{document}

\begin{frontmatter}


\title{Exploring Edge TPU for\\ Network Intrusion Detection in IoT}



\author[add1,add2]{Seyedehfaezeh Hosseininoorbin \footnote{Corresponding author.
\newline{{\it E-mail address:} f.noorbin@uq.net.au (Faezeh Noorbin).}}}
\author[add1]{Siamak Layeghy}
\author[add1]{Mohanad Sarhan}
\author[add3]{Raja Jurdak}
\author[add1]{Marius Portmann}

\address[add1]{School of Information Technology and Electrical Engineering, The University of ~~~~~\\Queensland, Australia~~~~~~~~~~~~~~~~~~~~~~~~~~~~~~~~~~~~~~~~~~~~~~~~~~~~~~~~~~~~~~~~~~~~~~~~~~~\\}
\address[add2]{DATA61, Commonwealth Scientific and Industrial Research Organisation (CSIRO), \\Australia~~~~~~~~~~~~~~~~~~~~~~~~~~~~~~~~~~~~~~~~~~~~~~~~~~~~~~~~~~~~~~~~~~~~~~~~~~~~~~~~~~~~~~~~~~~\\}
\address[add3]{School of Computer Science, Queensland University of Technology, Australia~~~~~~~~~~}

\begin{abstract}
This paper explores Google's Edge TPU for implementing a practical network intrusion detection system (NIDS) at the edge of IoT, based on a deep learning approach.
%
%
While there are a significant number of related works that explore machine learning based NIDS for the IoT edge, they generally do not consider the issue of the required computational and energy resources.
The focus of this paper is the exploration of deep learning-based NIDS at the edge of IoT, and in particular the computational and energy efficiency. 
In particular, the paper studies Google's Edge TPU as a hardware platform, and considers the following three key metrics: computation (inference) time, energy efficiency and the traffic classification performance.
Various scaled model sizes of two major deep neural network architectures are used to  investigate these three metrics.
The performance of the Edge TPU-based implementation is compared with that of an energy efficient embedded CPU (ARM Cortex A53).  Our experimental evaluation shows some unexpected results, such as the fact that the CPU significantly outperforms the Edge TPU for small model sizes.

\end{abstract}


\end{frontmatter}


\section{Introduction}

Practical network intrusion detection systems (NIDSs) in the context of IoT edge computing differ from non resource-constrained network solutions with ample computational and energy resources.
Indeed, what makes a solution for NIDS at the edge of IoT different from NIDSs for general networks is the considerations for the computational, energy and other resources constraints.
However, most of the proposed NIDSs for the IoT edge are general NIDSs that are proposed without investigating the limitations concerning the resources on edge devices for running such applications.

This paper explores Google's \textit{Edge TPU} for implementing NIDS at the IoT edge by investigating the requirements of the computational and energy constraints for running Deep Learning (DL) based algorithms.
The Edge TPU platform is a purpose-built ASIC hardware accelerator recently developed by Google to run inference at the edge~\cite{edgetpu}. 
The hardware accelerators are created to enhance the computational capabilities of the systems and make it possible to implement DL algorithms on the edge devices~\cite{Wisultschew2019,RN830,eTPUvsInteli9}. 
The NIDS algorithms utilised in this exploration are selected from two major deep neural network architectures, the feed forward and convolutional neural networks.
%

The practicality of Edge TPU for implementing NIDS at the edge of IoT is investigated in three aspects, the computational cost / inference time, energy efficiency  and attack detection / classification performance.
These three directions are explored by scaling the structure of the neural network architectures, i.e. generating neural networks with different model sizes.
The models are categorised into small and large sizes and the effect of each group on the above three parameters is studied via a public IoT benchmark dataset~\cite{ton-iot}.
%

The implementations of the NIDS algorithms on another hardware platform, ARM Cortex-A53 which is a power-efficient 64-bit embedded CPU, are used to further investigate the practicality of an Edge TPU-based NIDS.
All the experiments run on Edge TPU are repeated for this platform and the results are compared. 
There are several findings, including the two following major points. First, Cortex-A53, surprisingly, is faster and more energy efficient than Edge TPU for very small model sizes. Secondly, the energy efficiency and inference time of the Edge TPU platform  significantly depends on whether the neural network model size is larger or smaller than the size of on-chip memory of the platform.

The rest of this paper is organised as follows. The related works for network intrusion detection systems in the field of IoT are reviewed in Section~\ref{lit}.
Section~\ref{Classification Architecture}
explains the two neural network architectures used in this study and how their structures are scaled.
Section~\ref{setup} presents the dataset and hardware platforms used in this paper, and describes our experiments.
The results of the experiments are presented and discussed in Section~\ref{result}, and Section~\ref{con} concludes the paper.

\section{Related Work}\label{lit}

While there are many previous studies proposing NIDSs for the IoT edge, only a few have included studying the computational and energy resources utilisation / limitation in their proposals.
Since the focus of this paper is the practical NIDS at the edge, which concerns resource utilisation / limitation, only previous works with similar directions are included in this section.


%

Almogren~\cite{Almogren2020} proposed an IDS for Edge-of-Things networks based on deep belief neural networks (DBN). The author explored the DBN structure to find the best accuracy by adding hidden units to Restricted Boltzmann Machine (RBM) layers. The proposed DBN-based approach used to detect normal traffic and 9 different attack types.
The maximum accuracy achieved on a single attack classification is 85.73\% on UNSW-NB15 dataset~\cite{unsw-nb15}.
%
While deep belief neural network achieve promising accuracy, RBM-based approaches are not efficient in terms of inference time. 

Otoum et al.~\cite{Otoum2019}, have studied the feasibility of ML solutions for IDS in wireless sensor networks. 
For this purpose, they compared the performance of two different IDSs, RBM clustered IDS and the Adaptive Supervised and Clustered Hybrid (ADH) IDS. While their both methods achieve a F1-score of $\sim$99\% on KDD Cup 99~\cite{kddcup99}, the inference times for both methods are very long, for RBM it is 1.62 seconds and for ADH it is 0.86 seconds.

Eskandari et al.~\cite{Eskandari2020} proposed Passban IDS, which is a machine learning (ML)-based anomaly detection approach that uses the ensemble isolation forest. 
The proposed method was implemented on a Raspberry Pi3 as an IoT gateway and used 24.68\% (54 MB) of main memory and 47.17\% of CPU load and achieved a F1-score of 97.25\% on their own dataset collected from IoT testbed set up. 
While their solution achieves high performance, considering the 2 Watts power usage of Raspberry Pi3 with almost half used by its CPU~\cite{powerA53}, the proposed method is very inefficient in terms of the energy consumption. 

Hafeez et al.~\cite{Hafeez2020} proposed IoT-KEEPER IDS which uses fuzzy C-Means clustering and fuzzy interpolation scheme to analyse network traffic using TCP/IP features, and targeted different attack types. 
The prototype implementation of the method on a Raspberry Pi3, achieved 0.95\% and 0.93\% F1- score for binary and multiclass classification, respectively, on a combination of YTY2018~\cite{yty2018} and MSI2017~\cite{msi2017} datasets. In prototype implementation, IoT-KEEPER used 12\% (121 MB) of available memory and 4\% of CPU load.

Jan et al.~\cite{Jan2019} proposed an SVM-based IDS to detect DDos attack.
The authors used different statistics of the packet arrival rate, such as Mean, Median, Max and their combination to analyse data.
In the binary classification between DDoS / non-DDoS, the method achieved 98.03\% accuracy on CIC-IDS2017 dataset~\cite{cicids2017}. To show the efficiency of their approach, the authors compared the elapsed CPU time to simulate their NIDS, 2.28 seconds, with those of three previously proposed algorithms. However, they only used simulations for the evaluation of their method and provided no information regarding the required resources (memory, CPU, and energy) in their method.


Soe et al.~\cite{Soe2020} proposed an ML-based classifier along with a feature selection algorithm to implement a lightweight IDS. The feature selection algorithm is used to reduce the number of features as well as computational cost.
They proposed using different features for different attack types, and reduced the size of their classifier to fit on Raspberry Pi3. They used the Bot-IoT dataset~\cite{bot-iot} for the evaluation of their IDS, and achieved a F1-score of 98.42\%, the CPU Time of 0.81 seconds, and a memory usage of 425 MB. 
While they have investigated the resource utilisation of their proposed NIDS, which shows their concern about the practicality of their method, their proposed method is tailored to their evaluation dataset.
In other words, using a single feature to identify a specific attack type across various benchmark dataset is not currently confirmed.

The above studies showed a promising potential for using ML based methods for implementing NIDSs on IoT edge devices. However, most of these methods are using shallow learning  techniques which 
requires hand engineered features not generalisable to other benchmark datasets and real-world network traffic.
Moreover, 
wherever information relating the resource utilisation of these methods are provided they indicate their corresponding solution is hardly feasible on most of the IoT edge devices.
The two deep learning-based methods~\cite{Almogren2020, Otoum2019}, on the other hands, report very long processing time and do not provide information relating the resource utilisation of their proposed method.
As such, exploring the resource utilisation of ML-based and specially DL-based NIDSs, for implementation on IoT edge devices, is a crucial task in this field which is addressed by this paper.



\section{Classification Architecture}\label{Classification Architecture}
Two types of neural network architectures along with structural variations have been used to explore the resource utilisation of the hardware platforms considered in this study. 
The first architecture is the feed forward, which is the most trivial architecture in many deep learning problems. The next architecture is the convolutional neural network that is commonly used in many  DL-based NIDSs~\cite{NIDSCNN0,NIDSCNN1,NIDSCNN2,NIDSCNN3,NIDSCNN4}.

This study examines the effect of the model size on the capabilities of the hardware platforms for implementing the NIDS, and changing the structure of neural networks is utilised to scale the model sizes.
The models with a size of 800~KiB or less are called small and models with a size larger than that are called large models.
This is the model size of the biggest neural network with the feed forward architecture, and beyond this value, we have only extended the CNN architecture.
%
%
%
The two neural network architectures along with their structural variations are explained in this section.

\subsection{Feed Forward Neural Network}
\begin{figure}[!t]
 \centerline{\includegraphics[width=6.5 cm]{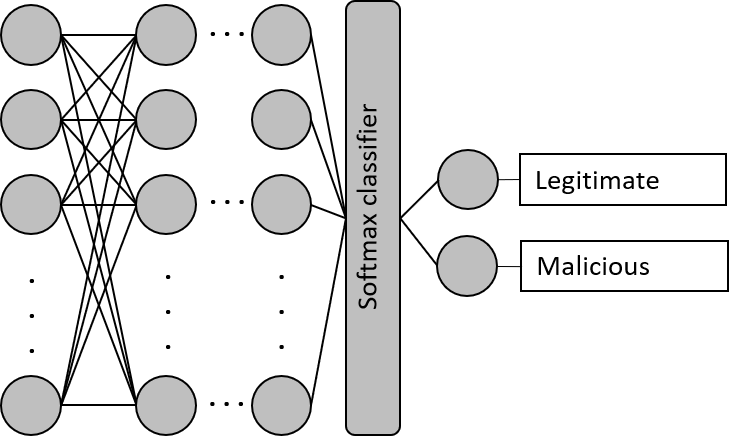}}
 \caption{The feed forward neural network architecture}
 \label{fig:densnet}
 \end{figure}

 Figure \ref{fig:densnet} shows an example feed forward neural network architecture used in this study. It 
 consists of a stack of fully connected layers including an input layer, various number ($M$) of hidden layers and the output layer. The number of nodes in the first layer equals number of data features / attributes, which is $39$ in the case of the dataset used in this paper. The hidden layers have $100$ nodes and the output layer is binary which distinguishes ``Legitimate" and ``Malicious". 
In order to scale the model size, 
number of hidden layers ($M$) is doubled in each step of the experiment, starting from 2 up to 64.
The resulting model sizes are less than 800~KiB referred to as small models in this study.
%
%
The activation function for all nodes is set to Rectified Linear Unit (ReLU) except the output layer where Softmax is used. The loss function for this model is Sparse Categorical Cross-entropy.

\subsection{Convolutional Neural Network}
\begin{figure}[!t]
 \centerline{\includegraphics[width=8.5 cm]{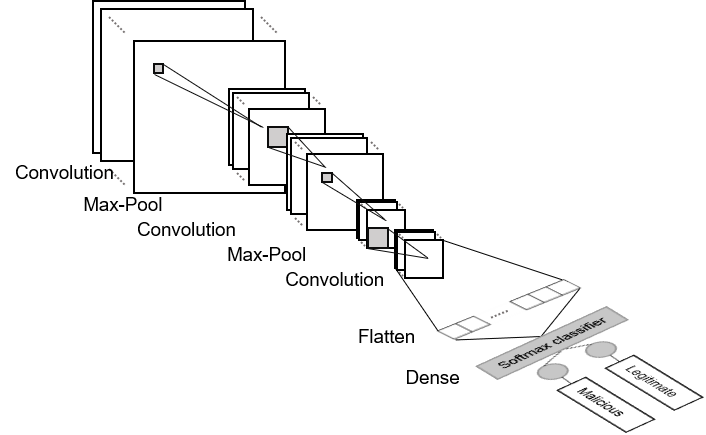}}
 \caption{The convolutional neural network (CNN) architecture}
 \label{fig:convnet}
 \end{figure}

The general form of the Convolutional Neural Network (CNN) architecture used in this paper is shown in Figure~\ref{fig:convnet}.
Two sets of experiments are conducted based on CNN architecture. 
In the first set of experiments that corresponds to small models, scaling the model size is achieved by changing the number of kernels and kernel size in convolution layers.
The CNN architectures in the first set of experiments generally include 3 convolution layers with $N$, $N/2$ and $N/2$ kernels of size $W\times W$ respectively, and max-pooling layers with $2 \times 2$ kernels between them. 
 The value of $N$ is doubled in each step of the experiments from 8 up to 256, with $W$ from $2$ to $5$.

 In the next set of experiments that corresponds to large models, while $N$ and $W$ are set to 256 and $5$ respectively, number of convolution layers is increased. The new  convolution layers are added only after the first convolution layer. The total number of convolution layers increases by the step of 3 in each experiment,  from 3 to 30. 
 %
%
The output of the last convolution layer is flattened and sent to the final dense layer with 2 nodes.
The ReLU activation function is used for all layers except the final, which is the Softmax function to generate either of the ``Legitimate" and ``Malicious" labels.

\section{Experimental Setup}\label{setup}
\subsection{Dataset}

A heterogeneous IoT dataset has been utilised to evaluate the proposed TPU-based NIDS. The ToN-IoT dataset~\cite{ton-iot} was released in 2019 by the Cyber Range and IoT Labs of UNSW Canberra based on a large-scale network testbed. It includes multiple data sources such as operating system logs, telemetry data of IoT/IIoT devices and IoT network traffic. 
The records containing labelled IoT network flow samples are utilised in this paper. It consists of nine attack scenarios including backdoor, DoS, Distributed DoS (DDoS), injection, Man In The Middle, password, ransomware, scanning and Cross-Site Scripting. 

The dataset is made up of mainly attack samples; 21,542,641 (96.44\%) and a low amount of benign samples; 796,380 (3.56\%). 
The Bro IDS was used to extract 44 features from IoT network traffic that make up the dataset.
There are 5 flow identifier features  that include timestamp (\textit{ts}), source IP address (\textit{src\_ip}), destination IP  Address (\textit{dst\_ip}), Source L4 Port (\textit{src\_port}) and destination L4 port  (\textit{dst\_port}).
The flow identifiers have been removed to avoid bias towards the attacking and victim end nodes in the classification experiments.

\subsection{Hardware Platforms}
This section explains the two hardware platforms used in the experiments of this paper, Google's Edge TPU and Raspberry Pi 3B+ that uses ARM Cortex-A53 as its processor.
 
 \subsubsection{Google Edge TPU}
\begin{figure}[!t]
 \centerline{\includegraphics[width=5 cm]{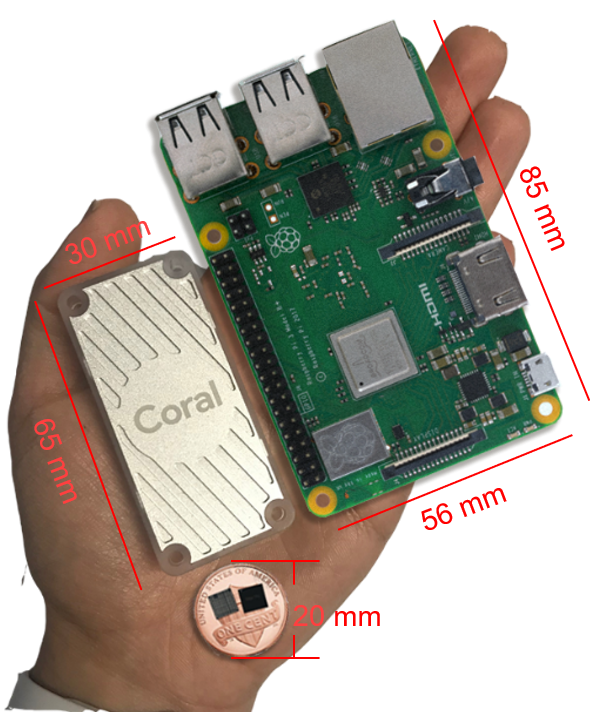}}
 \caption{Coral USB accelerator with Edge TPU and Raspberry Pi 3 model B+ platforms}
 \label{fig:Pi_TPU}
 \end{figure}

 The Edge TPU is a purpose-built ASIC hardware accelerator for machine learning (ML) applications with high performance and low energy footprint. In 2019, Google launched the Edge TPU with the aim of running machine learning inference at the edge~\cite{edgetpu}. 
 The TPU is based on a systolic array architecture, which allows highly efficient matrix multiplication at a massive scale. 
 All operations are limited to 8 bit integers, which increases both performance and energy efficiency~\cite{TPUarch}. 
 For the experiments presented in this paper, the Edge TPU Coral USB Accelerator~\cite{usbtpu} is utilised. Figure~\ref{fig:Pi_TPU} shows the Edge TPU Coral USB Accelerator (on the left), and the Edge TPU chip on top of a penny for reference.

\subsubsection{Raspberry Pi}
 The Raspberry Pi 3B+ is the second hardware platform used in this study. It is equipped with a 64-bit quad-core Arm Cortex-A53 CPU @1.4GHz with 1GB RAM. It is powered with a 5V-2.5A power supply running the Raspbian GNU/Linux OS. The Raspberry Pi 3B+ is can be connected to networks via gigabit Ethernet, integrated 802.11ac/n wireless LAN, and Bluetooth 4.2.
 %
 Figure~\ref{fig:Pi_TPU} (on the right), shows the credit card sized Raspberry Pi 3B+.

\section{Results}\label{result}
First, the results relating to the small models are presented, then  results relating to the large models are discussed, and finally both are compared. 

\subsection{Performance of Small Models}\label{small models}
The results relating small models includes three parts,
the trade-off between the model size and classification performance, the trade-off between model size and computational time, and the energy cost of performing an inference (classification). 

\subsubsection{Classification Performance}\label{Performance_F1}

 \begin{figure}[!b]
 \centerline{\includegraphics[width=\textwidth]{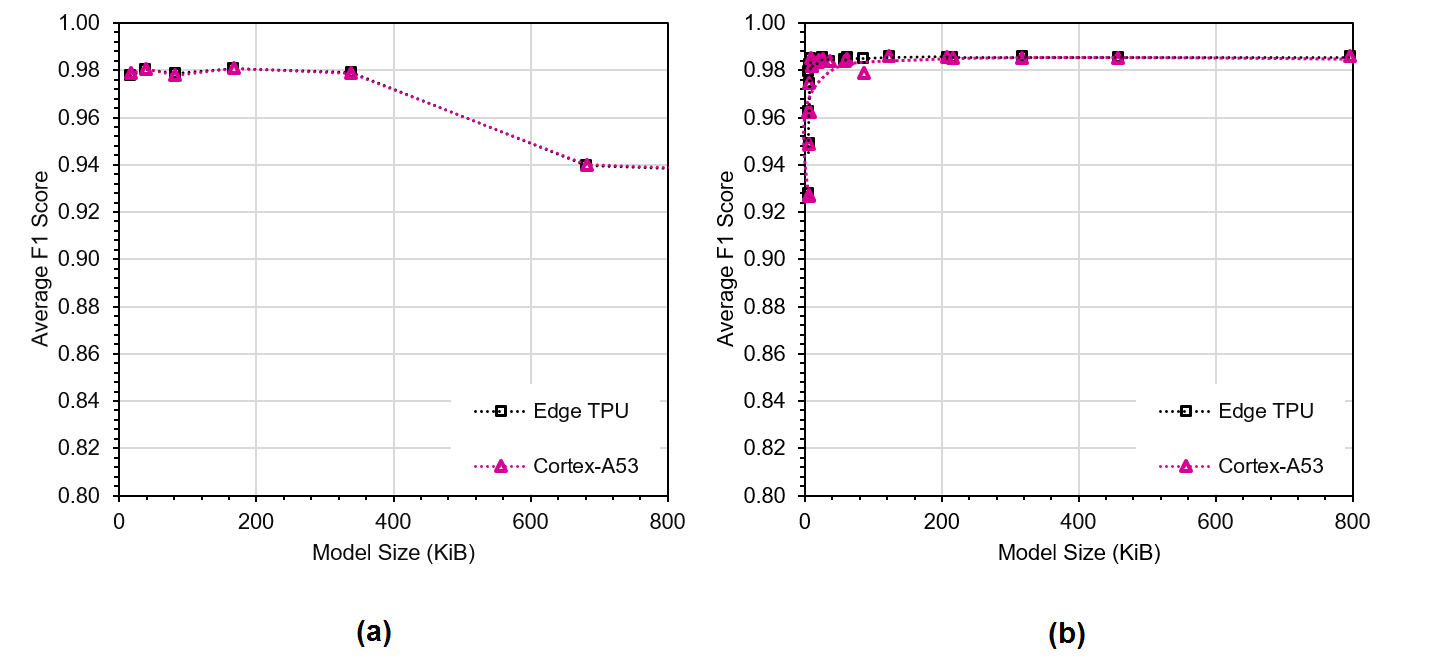}}
 \caption{The average F1-score versus model size for small models with (a) feed forward and (b) CNN architectures on Edge TPU and Cortex-A53}
 \label{fig:tiny_F1}
 \end{figure}

Figures~\ref{fig:tiny_F1} (a) and (b) show the classification performance of the feed-forward and convolutional architectures respectively as a function of model size on two hardware platforms. As seen, the difference between the two hardware platforms is not significant. 
The horizontal axis is model size in KiB and the vertical axis is the F1-score as a metric of the classification performance.

The variation of the model size for the feed forward models (Figure~\ref{fig:tiny_F1} (a)) is created by 
varying the number of the fully connected  hidden layers ($M$).
%
On both hardware platforms, the F1-score is increasing for the model sizes up to 338~KiB where it starts to drop. The highest achieved value of F1-score  is 0.98, which relates to values of $M$ between 2 and 32.

The variation of the model size for the CNN architecture 
(Figure~\ref{fig:tiny_F1} (b)) is created by changing the number of kernels $N$ and kernel sizes $W\times W$.
%
The model size of 5.2~KiB, corresponding to ($N$=$8$, $W$=$2$), has achieved the lowest value of the F1-score in all experiments.
%
Initially, the F1-score rapidly increases with the model size, but it is very sensitive to model parameters, swinging between 0.93 and 0.98 for model sizes up to 6.3~KiB. 
%
By increasing the model size up to 7.4~KiB, corresponding to ($N$=$16$, $W$=$3$), the F1-score becomes stable around 0.98, where it remains fix for all the larger model sizes. 
%
%
This is in contrast with the feed forward models (Figure~\ref{fig:tiny_F1} (a)) in which increasing the model sizes beyond 338~KiB reduces the F1-scores.


\subsubsection{Computational Time}\label{Performance_Time}
\begin{figure}[!b]
 \centerline{\includegraphics[width=\textwidth]{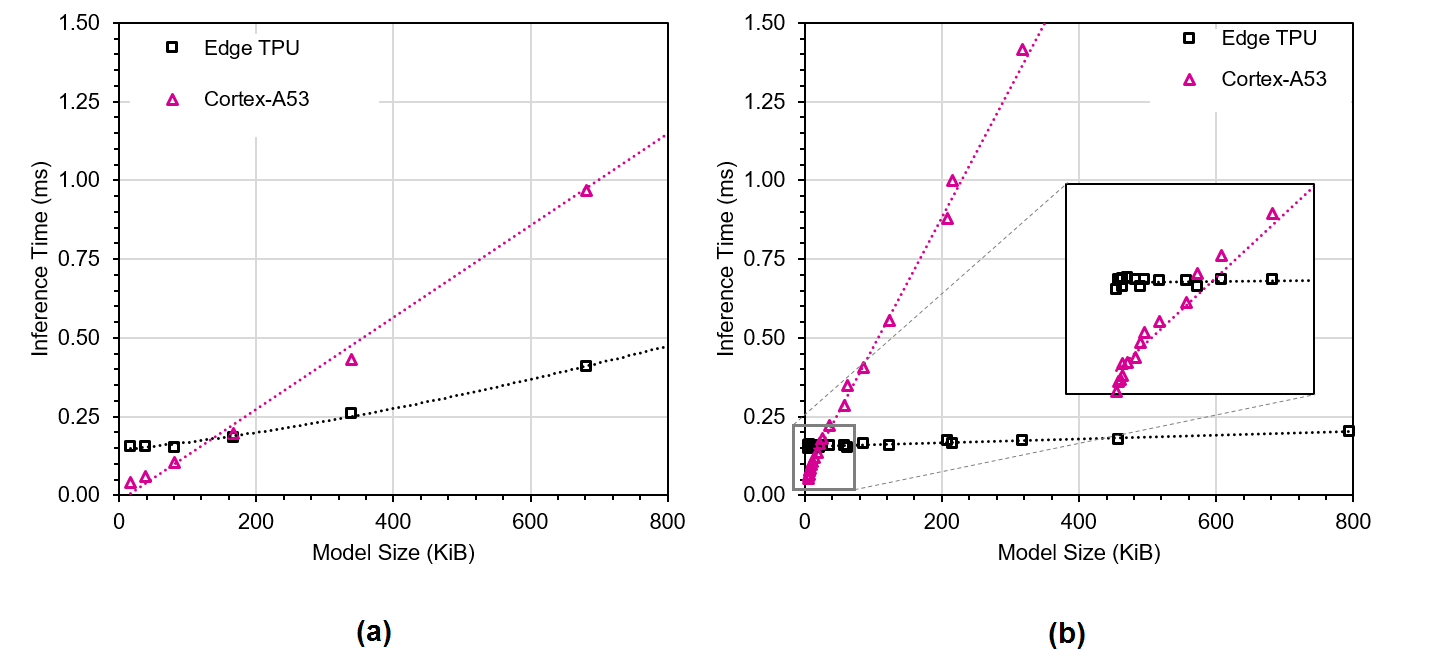}}
 \caption{The inference time versus model size for small models with (a) feed forward and (b) CNN architectures on Edge TPU and Cortex-A53}
 \label{fig:tiny_time}
 \end{figure}

Figure~\ref{fig:tiny_time} shows the inference time of the feed-forward (on the left) and CNN (on the left) architectures respectively as a function of model size, on both hardware platforms. 
%
The horizontal axis indicates model size in KiB and the vertical axis indicates the time needed to run a single inference using the corresponding architecture in Milliseconds. 

Surprisingly, Cortex-A53 is faster than Edge TPU for  very small model sizes. 
In the case of the feed forward architecture (Figure \ref{fig:tiny_time} (a)), the advantage of  
Cortex-A53 continues for model sizes up to $136$~KiB, but for the case of CNN architecture (Figure \ref{fig:tiny_time} (b)) this advantage only continues for model sizes up to $25$~KiB.

 Edge TPU outperforms Cortex-A53 in both neural network architectures and performs faster for all larger model sizes. 
It is also noticeable that the growth of the advantage of Edge TPU over Cortex-A53 is much faster in the case of the CNN architecture than for the feed forward architecture.


\subsubsection{Energy Efficiency}
While the energy needed to run a set of operations on a hardware platform is related to the time taken to perform the operations, since different hardware platforms have a different power consumption, the energy efficiency is independently investigated.
The energy efficiency is defined in this context as the number of inferences that can be performed with an energy budget of $1$~Milli-Joule~(mJ).

 \begin{figure}[!b]
 \centerline{\includegraphics[width=\textwidth]{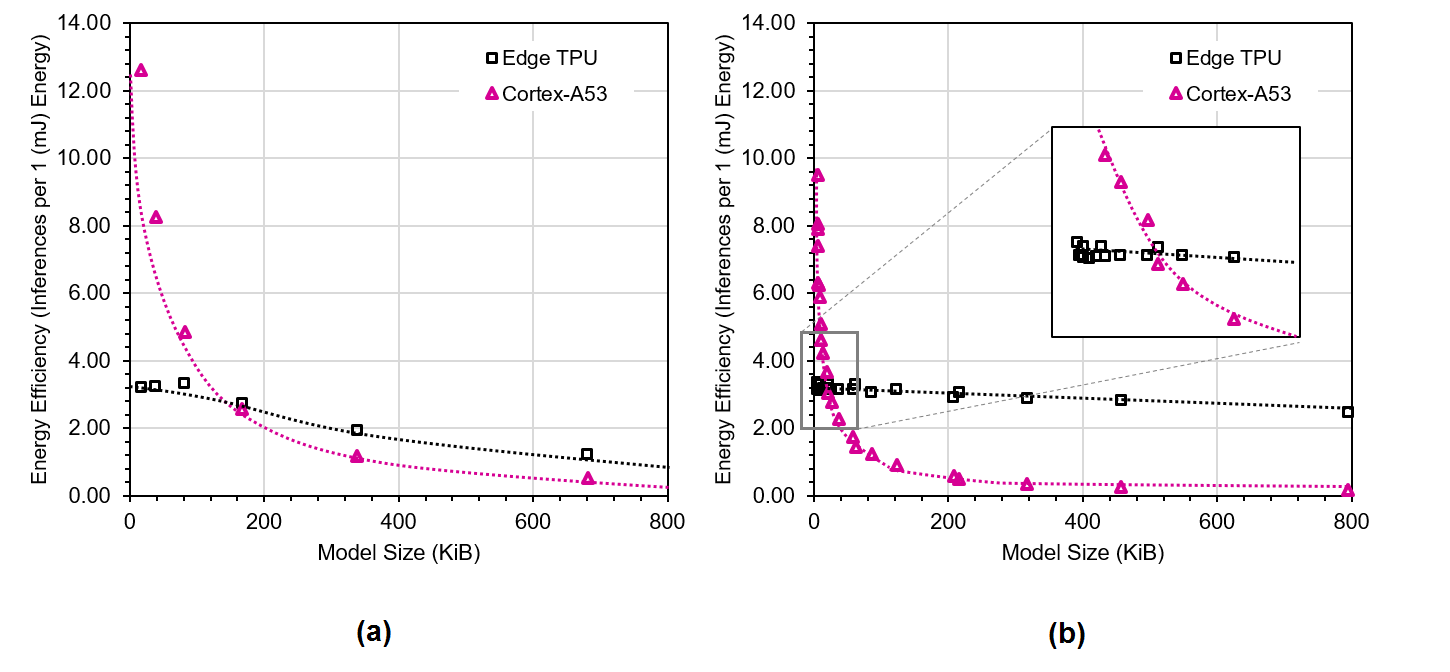}}
 \caption{Energy efficiency versus model size for small models with (a) feed forward and (b) CNN architectures on Edge TPU and Cortex-A53}
 \label{fig:tiny_energy}
 \end{figure}

Figure~\ref{fig:tiny_energy} (a) and (b) show the energy efficiency for the feed-forward and convolutional architectures respectively, as a function of the model size, on both hardware platforms.
The horizontal axis indicates model sizes in KiB and the vertical axis indicates the energy efficiency.
%
The Cortex-A53 platform has the initial advantage, i.e. it is more energy efficient than the Edge TPU for smaller model sizes. The superiority of Cortex-A53 continues for all model sizes up to 136~KiB and 25~KiB for the feed forward and CNN architectures respectively.

Again, after the initial advantage of Cortex-A53, the Edge TPU leads the energy efficiency race for all the larger model sizes. 
It is also noticeable that the Edge TPU's lead over Cortex-A53 is bigger when using the CNN architecture.
In other words, the Edge TPU is more energy efficient when using CNN architecture than using the feed forward architecture, for the same model size.

 \begin{figure}[!b]
 \centerline{\includegraphics[width=8.5 cm]{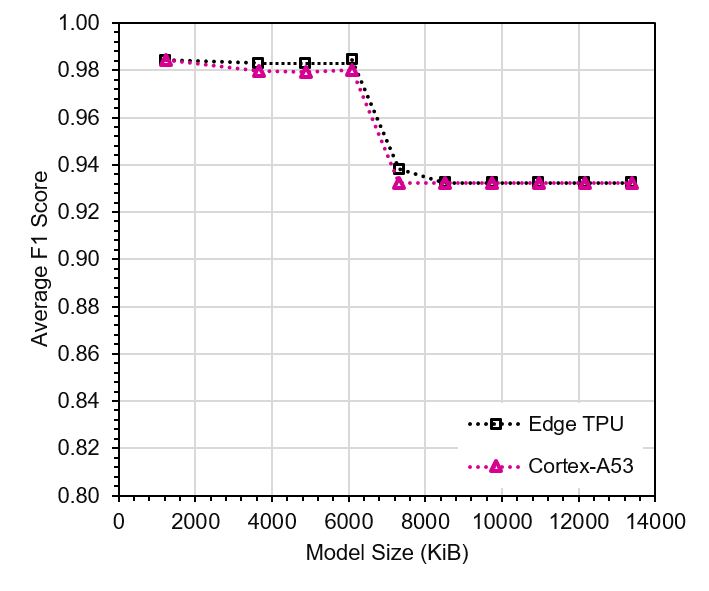}}
 \caption{The average F1-score versus model size for large models with CNN architecture on Edge TPU and Cortex-A53}
 \label{fig:mega_F1_conv}
 \end{figure}

\subsection{Performance of Large Models}\label{large models}
In addition to comparisons conducted for the small models including the classification performance, computational time and energy efficiency, the memory usage of the large models is also discussed in the following. As mentioned before, the large models are only investigated using the CNN architecture where the increase of the number of convolution layers from 3 to 30 by step of 3 is used to scale the model size ($N$=256 and $W$=5 are fixed).
 
\subsubsection{Classification Performance}


 Figure~\ref{fig:mega_F1_conv} shows the classification performance for the large models. 
 The F1-score is $\sim$0.984 for model sizes up to $6000$KiB, corresponding to 12 convolution layers, and drops for models beyond that to $\sim$0.932 on both hardware platforms. 
 While this performance degradation is a clear indication for not using larger models for this problem, large models have been investigated in this paper to study other factors such as memory usage, inference time and energy efficiency of the hardware platforms in the broader range.


\subsubsection{Computational Time}

\begin{figure}[!b]
 \centerline{\includegraphics[width=8.5 cm]{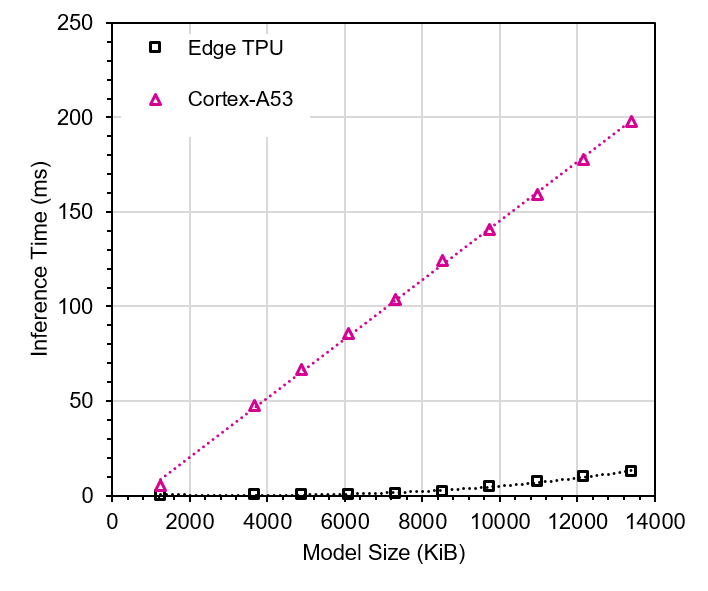}}
 \caption{The inference time versus model size for large models with  CNN architecture on Edge TPU and Cortex-A53}
 \label{fig:mega_time_conv}
 \end{figure}

Figure~\ref{fig:mega_time_conv} illustrates the inference time for large CNN models on both hardware platforms, as a function of the model size. As can be seen, the inference time of similar model sizes on Edge TPU are much smaller than on Cortex-A53.
The other phenomenon observed in the results is the bi-modal behaviour of the Edge TPU platform. While the slope of the linear curve (inference time per model size) is very small for model sizes up to 8000KiB, the slope becomes much bigger after this value.
The bi-modal behaviour of Edge TPU is connected to its on-chip memory usage, as also observed in \cite{Workshop2021}. As such, the memory usage of the CNN architecture with model sizes in the range of these behavioural modes is investigated in the next sub-section.

%

\subsubsection{Memory Usage}
\begin{figure}[!b]
 \centerline{\includegraphics[width=8.5 cm]{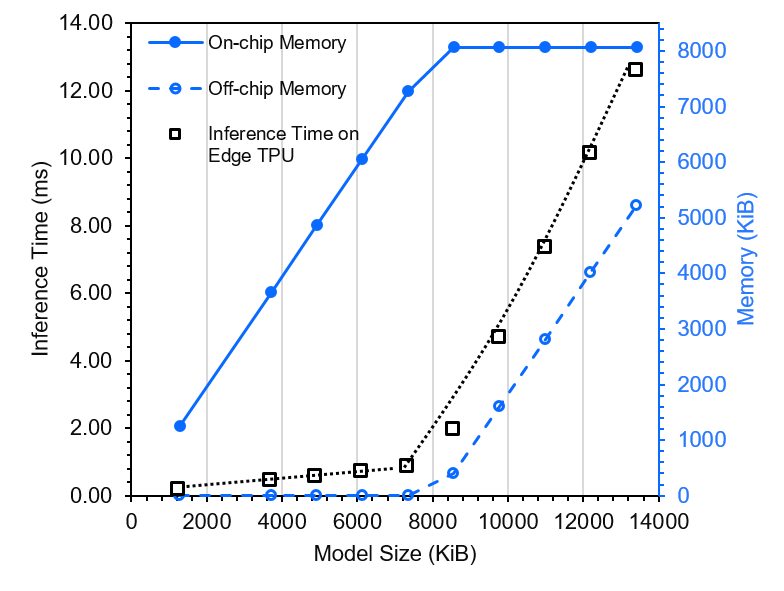}}
 \caption{The memory usage (right vertical axis) and inference time (left vertical axis) versus  model size for large models with CNN architecture on Edge TPU and Cortex-A53}
 \label{fig:mega_memory_conv}
 \end{figure}

Figure ~\ref{fig:mega_memory_conv} illustrates the memory usage of the Edge TPU platform throughout its bi-modal behaviour range. The figure is created by running CNN architectures with model sizes in both modes, and recording the corresponding inference time, on-chip and off-chip memory usage.  
The horizontal axis is the model size in KiB, the left vertical axis indicates the inference time in Milliseconds and the right vertical axis indicates the utilised memory in KiB.

The curve of the inference time consists of two linear curves. The first curve starts from a model size of about 1000~kiB and finishes at a model size of 8000~kiB. This is the beginning of the second linear curve with a different slope which continues to model sizes up to 14000~KiB (and possibly beyond).
Similar trends can be seen in the case of the other two curves, the on-chip and off-chip memory usages. 
The turning point of all three curves is at the model size of 8000~KiB.
Investigating the on-chip memory curve makes this very clear. The on-chip memory curve increase up to 8000~KiB with a slope of $\sim$1, which means the on-chip memory usage is equal to the model size. In other words, the entire CNN model is loaded into the on-chip memory, and the off-chip memory usage is near zero.

Once the model sizes get bigger than 8000~KiB, the inference time significantly increases and the second behavioural mode appears. The on-chip memory usage remains at 8000~KiB for all larger model sizes, which is the size of the on-chip memory of the Edge TPU platform.
Simultaneously, the off-chip memory usage starts growing with the increase of the model size, which means part of the CNN models are being loaded into the off-chip memory. The inference time increases as more model parts are loaded into the off-chip memory.




\subsubsection{Energy Efficiency}
 
 

 \begin{figure}[!b]
 \centerline{\includegraphics[width=8.5 cm]{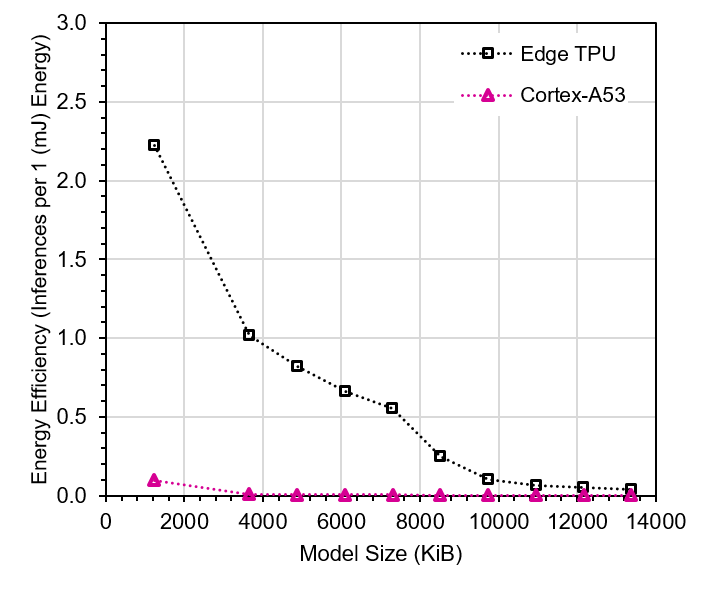}}
 \caption{Energy efficiency versus model size for large models with CNN architecture on Edge TPU and Cortex-A53}
 \label{fig:mega_energy_conv}
 \end{figure}
  
Figure~\ref{fig:mega_energy_conv} shows the energy efficiency of the two considered hardware platforms for large models as a function of the model size. 
The first observation is the decline of the energy efficiency with the increase of the model size on both hardware platforms, which is not a surprise.
The other observation is the superiority of the Edge TPU over  Cortex-A53 in terms of the energy efficiency, for all large models.
While Edge TPU leads over Cortex-A53 for all large model sizes, its advantage is much larger for model sizes below 8000~KiB. 
This is another manifestation of the Edge TPU's bi-modal behaviour.
For  model sizes beyond 8000~KiB, the energy efficiency of the two platforms converges and becomes less efficient.

 %
%


\subsection{Summary of The Small and Large Models}
\begin{figure}[!b]
 \centerline{\includegraphics[width=8.5 cm]{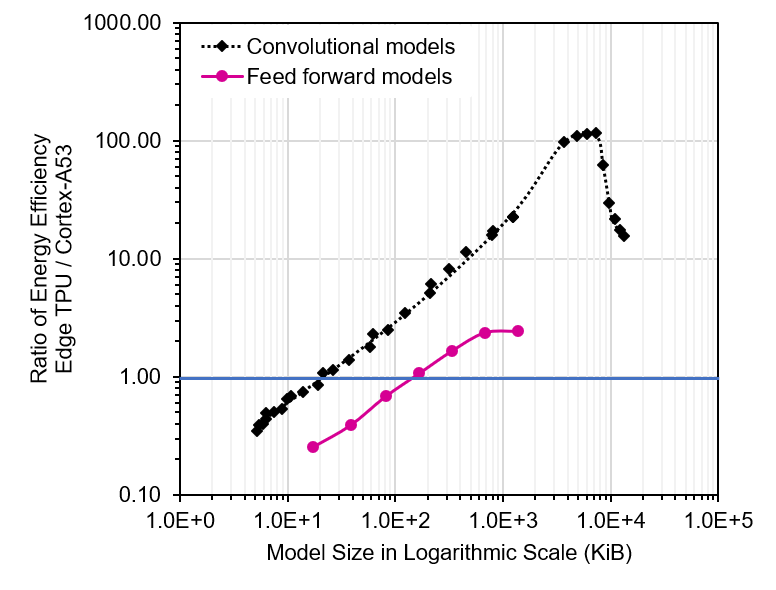}}
 \caption{The ratio for energy efficiency of Edge TPU over Cortex-A53 in terms of model size for both small and large models using the CNN and feed forward architectures}
 \label{fig:Sum_ratio_misc}
 \end{figure}

In order to have an overall view of the  performance comparison for both small and large models, the results form Section \ref{small models} and Section \ref{large models} are summarised  in Figures~\ref{fig:Sum_ratio_misc}.
The figure shows the ratio of the energy efficiency of the Edge TPU over Cortex-A53 for both neural network architectures, in terms of the model size, including both small and large models. 
The horizontal axis indicates the model sizes in logarithmic scale, and the vertical axis indicates the ratio of the energy efficiency of Edge TPU over Cortex-A53, also in logarithmic scale.

The first observation is that using the CNN architecture on the Edge TPU platform is more energy efficient than using the feed forward architecture, independent of the model size.  
The CNN architecture is 8 to 10 times more energy efficient than feed forward models when used on the Edge TPU platform.

The next observation is the superiority of Cortex-A53 over Edge TPU for both neural network architectures, when using very small model sizes. The curves corresponding to both architectures start at a ratio below 1 (the horizontal blue line), which means Cortex-A53 is more efficient than Edge TPU.
However, by increasing the model sizes, both curves soon pass the blue horizontal line, meaning the Edge TPU becomes more energy efficient. 
Again, the CNN architecture passes the blue line for much smaller model sizes, which means using the CNN architecture is more energy efficient on Edge TPU for much smaller model sizes than for the feed forward architecture.

The last observation is the bi-modal behaviour of the Edge TPU. As seen, the advantage of CNN over the feed forward model increases when we  increase the model size. However, once Edge TPU starts using the off-chip memory, i.e. about a model sizes of 8000~KiB, increasing the model size further reduces the relative advantage of the Edge TPU.


   



\section{Conclusion}\label{con}
This paper explores Google's  Edge TPU for the implementation of a practical deep learning-based NIDS at IoT edge.
%
The Edge TPU-based NIDS is evaluated on a public IoT benchmark dataset.
The detection performance, energy efficiency and inference time are the three aspects investigated using this dataset, for varying neural network models sizes.
Two neural network architectures are considered in this exploration, feed forward and convolutional neural networks.

In order to further investigate the practicality of an Edge TPU-based NIDS in IoT, and for comparison,  all the created neural network models are also implemented on the ARM Cortex-A53, a power-efficient 64-bit embedded CPU.
The result of our comparison indicate that Cortex-A53, somewhat surprisingly, is faster and more energy efficient than the Edge TPU for very small neural network models. 
%

Our results also show that, beyond small model sizes, the Edge TPU platform is always superior to Cortex-A53, for both considered neural network architectures.
We further observed that the advantage of Edge TPU over Cortex-A53 declines once the model size reaches the size of the on-chip memory of the Edge TPU platform. This bi-modal behaviour of Edge TPU, which is also reported in \cite{Workshop2021}, is not currently documented by Google.
This clearly shows the importance of considering the Edge TPU's on-chip memory size and the neural network model size,  when designing deep learning-based systems for IoT/edge applications, and for making a decision whether the Edge TPU is the right hardware platform for such an application. 
The final finding of these experiments is that the CNN architecture on the Edge TPU platform is more energy efficient than feed forward models, for all model sizes.

 \section*{}

 \bibliography{References} 
\end{document}